\documentclass[runningheads]{llncs}

\usepackage[hyphens]{url}

\usepackage{graphicx}
\usepackage{xcolor}
\usepackage{multirow}
\usepackage{tikz,hyperref}
\usepackage{wasysym} 
\usepackage{multirow} 

\usepackage{amssymb,amsfonts}
\usepackage{textgreek} 

\begin{document}



\title{Securing IIoT using Defence-in-Depth: Towards an End-to-End Secure Industry 4.0}
\titlerunning{Securing IIoT using Defence-in-Depth}


\author{Aintzane Mosteiro-Sanchez\inst{1,2} \and Marc Barcelo\inst{1} \and Jasone Astorga\inst{2} \and Aitor Urbieta\textbf{\inst{1}}}

\authorrunning{A. Mosteiro-Sanchez et al.}

\institute{Ikerlan Technology Research Centre, Basque Research and Technology Alliance (BRTA). P.º J.M. Arizmendiarrieta, 2. 20500 Arrasate/Mondragón, Spain \\
\email{\{amosteiro,mbarcelo,aurbieta\}@ikerlan.es} \and
Department of Communications Engineering, Faculty of Engineering, University of the Basque Country UPV/EHU,Alameda Urquijo s/n, 48013 Bilbao, Spain
\email{jasone.astorga@ehu.eus}}

 \maketitle

\begin{abstract}
Industry 4.0 uses a subset of the IoT, named Industrial IoT (IIoT), to achieve connectivity, interoperability, and decentralization. The deployment of industrial networks rarely considers security by design, but this becomes imperative in smart manufacturing as connectivity increases. The combination of OT and IT infrastructures in Industry 4.0 adds new security threats beyond those of traditional industrial networks. Defence-in-Depth (DiD) strategies tackle the complexity of this problem by providing multiple defense layers, each of these focusing on a particular set of threats. Additionally, the strict requirements of IIoT networks demand lightweight encryption algorithms. Nevertheless, these ciphers must provide E2E (End-to-End) security, as data passes through intermediate entities or middleboxes before reaching their destination. If compromised, middleboxes could expose vulnerable information to potential attackers if it is not encrypted throughout this path. This paper presents an analysis of the most relevant security strategies in Industry 4.0, focusing primarily on DiD. With these in mind, it proposes a combination of DiD, an encryption algorithm called Attribute-Based-Encryption (ABE), and object security (i.e., OSCORE) to get an E2E security approach. This analysis is a critical first step to developing more complex and lightweight security frameworks suitable for Industry 4.0.

\end{abstract}


\section{Introduction} \label{S:1}
In recent years, \textit{IoT} has become a popular term used in many areas. Although there is no official definition, several attempts have been made in this direction~\cite{ETSI-EuropeanTelecommunicationsStandardsInstitute2013Machine-to-MachineRequirements}~\cite{ITU2019InternetInitiative}~\cite{Minerva2015TowardsIoT}, which usually describe the IoT as a set of connected devices able to process, send or receive data, with or without an Internet connection. This has transformed the way people and machines communicate and interact with each other. Nowadays, the IoT revolution has reached the industry, leading to the fourth industrial revolution~\cite{Rojko2017IndustryOverview}, or Industry 4.0. \par

\textit{Industry 4.0} is a concept coined by the German Government~\cite{BundesministeriumfurBildungundForschung2017IndustrieMorgen} and presented in the Hannover Messe 2011. It aims to produce higher quality products and reduce production costs through Industrial IoT (IIoT), among other key enabling technologies. IIoT is a subset of the IoT applied to industry, and the evolution of industrial communications~\cite{Sisinni2018IndustrialDirections}. It increases connectivity, interoperability, and decentralization. IIoT devices collect the exchanged information en masse, which has increased in volume, variety, and complexity~\cite{Tao2018Data-drivenManufacturing}. This means that the data volume IIoT devices need to manage tends to be much higher than typical IoT applications. Various researchers have analysed the properties and constraints of IoT and IIoT~\cite{Sisinni2018IndustrialDirections}~\cite{Atzori:2010:ITS:1862461.1862541}~\cite{Zhou2019TheSolved}. They are summarised in Table \ref{tab:1_IIoTVSIoT}, where \textbf{!} symbolizes that it may not exist in every Industry 4.0 environment. Features like interdependence are of especial relevance in industry since it implies that even if these constrained features only affect a few nodes in the network, they can potentially impact the entire system. For example, an uncontrolled alteration in the read of a sensor can affect the actuators and control system, risking the availability of the entire system.\par

\begin{table}[!ht]
\caption{\label{tab:1_IIoTVSIoT} Feature comparison between IoT and IIoT.}
\centering
\resizebox{60mm}{!}{%
\begin{tabular}{l|c|c|}
\cline{2-3}
                                                     & \textbf{IoT}  & \textbf{IIoT} \\ \hline
\multicolumn{1}{|l|}{\textbf{Battery Limitation}}    & {\checkmark} & {!} \\ \hline
\multicolumn{1}{|l|}{\textbf{Computing Limitation}}  & {\checkmark } & {\checkmark}  \\ \hline
\multicolumn{1}{|l|}{\textbf{Sleep-Mode}}            & {\checkmark } & {!} \\ \hline
\multicolumn{1}{|l|}{\textbf{Interdependance}}       & {\checkmark } & {\checkmark}  \\ \hline
\multicolumn{1}{|l|}{\textbf{Heterogeneity}}         & {\checkmark} & {\checkmark}  \\ \hline
\multicolumn{1}{|l|}{\textbf{Structured Nodes}}      & {$\times$} & {\checkmark}  \\ \hline
\multicolumn{1}{|l|}{\textbf{Scalability}}           & {\checkmark } & {\checkmark}  \\ \hline
\multicolumn{1}{|l|}{\textbf{Interoperability}}      & {\checkmark } & {\checkmark}  \\ \hline
\multicolumn{1}{|l|}{\textbf{Very High Data Volume}} & {$\times$} & {\checkmark}  \\ \hline
\end{tabular}
}
\end{table}

Because of the constrained nature of IIoT devices, sometimes data processing is carried out in edge devices, or the Cloud~\cite{Aazam2018Deploying4.0}~\cite{WU2017316}, and even floor scheduling and condition-based maintenance~\cite{Mourtzis2018AMaintenance}. Thus, wireless communications are increasingly common in industrial environments, using protocols such as Zigbee, WirelessHART, Trusted Wireless, WiFi, or Bluetooth~\cite{INCIBE-CERT2017CiberseguridadIndustriales}. The application layer protocols running on top of them should be lightweight and address the constrained nature of IIoT devices. Therefore, protocols and security solutions designed for the IoT may suit IIoT. In this context, the lightweight protocols proposed by the IETF Working Group, CoRE~\cite{CORE2020Constrainedcore}, may be of interest in smart manufacturing.\par
Industry 4.0 networks have to deal with IIoT devices and enhanced connectivity and interoperability. Additionally, due to the long life span of the Operational Technology (OT) devices, legacy-related issues must be considered too. Examples of these issues are limitations in the communications buffer, the lack of security patches~\cite{book:1375056}, problems to implement authentication~\cite{EuropeanUnionAgencyforNetworkandInformationSecurityENISA2019INDUSTRYRECOMMENDATIONS}, difficulties to upgrade legacy systems without affecting system availability~\cite{Brachmann2012End-to-endThings} or interoperability issues with newer systems~\cite{8012471}. Not addressing the limitations of legacy devices might cause various incidents, e.g., safety violations, monetary losses, or information theft. The complexity of industrial networks makes their security complex too, so strategies like Defence-in-Depth (DiD) are used. This security strategy creates different security layers, with the idea that if attackers enter the system, the security measures will hinder them for long enough to be detected~\cite{Nguyen2017MicrosoftGuide}. Monitoring and authentication systems that detect these attackers are established throughout all the layers. In this way, even in the case of legacy devices that cannot authenticate users, they can form part of a network that has already been authenticated and whose traffic is being monitored. Therefore, the chances of an unauthorized user accessing that legacy system would be significantly reduced. Furthermore, the hierarchical structure of the industrial domain~\cite{Granzer2011SecuritySystems} facilitates establishing the layered approach of DiD. \par

Although industrial systems are still very hierarchical, Industry 4.0 architectures are becoming decentralized systems~\cite{MOGHADDAM2018215}, in which messages go through proxies, gateways, and other middleboxes to save bandwidth and memory or perform protocol translation operations~\cite{rfc8576}. These middleboxes provide scalability, efficiency, and interoperability among nodes. However, they have full access to the relayed data, even if communications have been protected with Transport Layer Security (TLS). Since TLS only protects the communication channel and not the message itself, this might cause security incidents if the middleboxes are compromised. In this case, TLS is not enough. Instead, additional End-To-End (E2E) security mechanisms are required to guarantee that data is not exposed to third parties. More about this will be explained in Section \ref{S:4.1}.\par

With this in mind, the purpose of this document is to study the security measures available for industry 4.0. It also considers the need for E2E security and how to combine it with a DiD strategy. To this end, we review the applicability in industrial environments of security techniques like object security and encryption ciphers like Attribute-Based-Encryption (ABE). Concepts such as DiD, encryption, and object security are extensively explained throughout the document.\par
The remaining paper is structured as follows: Section \ref{S:2} presents an overview OF industrial security, points out the most relevant Industry 4.0 security requirements, and provides security best practices for such scenarios. Section \ref{S:3} introduces the goals for a DiD strategy and explains security measures in each of the layers to comply with those objectives. Section \ref{S:4} analyses the need and implications of using encryption in manufacturing and how it can be used to obtain E2E security. Section \ref{S:5} and Section \ref{S:6} introduce object security (i.e., OSCORE) and ABE and discuss their applicability in Industry 4.0 scenarios. Finally, Section \ref{S:7} highlights the most important insights and concludes the paper. \par
 
\section{Security in Industry 4.0: A general approach} \label{S:2}
The particular features of IIoT must be considered when designing the industrial security system. Traits like battery and computing limitations will restrict the possible security solutions. Besides, Industry 4.0 uses other enabling technologies that go beyond IIoT, which will also significantly affect security. In the case of manufacturing, systems are complex structures formed by Information Technology (IT) and OT networks. IT networks refer to the technologies used for information processing and telecommunications equipment. OT networks are related to industrial equipment responsible for monitoring and controlling physical devices. Effective security architectures should be included since the system design stage and reviewed often~\cite{Tuptuk2018SecuritySystems}. They should also take into account the growing connectivity of OT networks, which makes them resemble IT networks more than ever~\cite{Harp2015ITDivide}, while still needing to remain separated, e.g., by keeping IT and OT infrastructures separate using Next-Generation Firewalls (NGFWs). These Firewalls offer application-level inspection, providing greater control over what enters and leaves the network, improving security, and facilitating updates~\cite{Thomason2012ImprovingDevices}. In terms of security, OT and IT have different priorities. IT follows the CIA triad, i.e., confidentiality, integrity, and availability, in that order. Meanwhile, this priority order is changed in the OT domain~\cite{Heer2016Secure4.0}. This is summarised in Table \ref{tab:2_SecRequirements}. \par

\begin{table}[!ht]
\caption{\label{tab:2_SecRequirements} Prioritisation of security requirements for IT and OT networks~\cite{Heer2016Secure4.0}.}
\centering
\resizebox{65mm}{!}{
\begin{tabular}{|c|c|c|}
\hline
\textbf{Priority Level} & \textbf{OT}    & \textbf{IT}    \\ \hline
\textbf{1}             & Availability   & Confidentiality \\ \hline
\textbf{2}             & Integrity      & Integrity      \\ \hline
\textbf{3}             & Confidentiality & Availability   \\ \hline
\end{tabular}
}
\end{table}

Differences between OT and IT have been widely studied in the literature and are not the focus of this paper. Still, addressing them is essential to understand why traditional IT security approaches cannot be directly applied to OT networks. Their most relevant traits from a security point of view are shown in Table \ref{tab:3_ITOTDiff}, which summarises the analysis presented in~\cite{Stouffer2015GuideSecurity}. It is particularly relevant to highlight the strict latency requirements, the need for a fault-tolerant design, or the much longer lifetime of OT systems compared to IT systems. These particularities should be considered when adapting existing IT solutions to the OT environment. For instance, DiD strategies. \par

\begin{table}[!ht]
\caption{\label{tab:3_ITOTDiff} Summary of OT and IT networks differences~\cite{Stouffer2015GuideSecurity}.}
\centering
\resizebox{110mm}{!}{%
\begin{tabular}{c|c|c|}
\cline{2-3}
\textbf{}                                                                                          & \textbf{OT}                                                                                          & \textbf{IT}                                                                                           \\ \hline
\multicolumn{1}{|c|}{\textbf{\begin{tabular}[c]{@{}c@{}}Performance\\  requirements\end{tabular}}} & \begin{tabular}[c]{@{}c@{}}Real-Time\\ Delays unacceptable\end{tabular}                          & \begin{tabular}[c]{@{}c@{}}No Real-Time\\ Delays acceptable\end{tabular}                          \\ \hline
\multicolumn{1}{|c|}{\textbf{Fault-Tolerance}}                                                     & Essential                                                                                            & Not important                                                                                         \\ \hline
\multicolumn{1}{|c|}{\textbf{Updates}}                                                             & \begin{tabular}[c]{@{}c@{}}Should first be implemented \\ in a controlled environment\end{tabular}     & \begin{tabular}[c]{@{}c@{}}Updates are \\ straightforward\end{tabular}                                \\ \hline
\multicolumn{1}{|c|}{\textbf{Communications}}                                                      & \begin{tabular}[c]{@{}c@{}}Proprietary protocols\\ Wired and Wireless\\ Complex Networks\end{tabular} & \begin{tabular}[c]{@{}c@{}}Standard protocols\\ Wired networks\\ IT networking practices\end{tabular} \\ \hline
\multicolumn{1}{|c|}{\textbf{Lifetime}}                                                            & 10-15 years                                                                                          & 3-5 years                                                                                             \\ \hline
\multicolumn{1}{|l|}{\textbf{Device Location}}                                                     & May be remote and isolated                                                                                  & Local and easy to access                                                                              \\ \hline
\end{tabular}%
}
\end{table}

\subsection{General Security Recommendations} \label{S:2.1}
Unfortunately, poor security practices have been discovered in industrial networks, like those emulated in ~\cite{Hilt2020CaughtThreats}. These security flaws mainly affect small and medium enterprises, which do not always have the required knowledge or resources to invest in strong security mechanisms and equipment~\cite{Schroder2016TheEnterprises}. However, we have determined that it is important to follow at least the following recommendations: \par

\begin{itemize}
   \item \textbf{Keep software up-to-date:} Enterprises sometimes use hardware with known vulnerabilities, e.g., Allen-Bradley's MicroLogix~\cite{brumaghin_vulnerability_nodate}~\cite{noauthor_multiples_2017} or Siemens Simatic~\cite{Siemens2019SSA-232418:Families}. To patch them, it is recommended to apply the security updates provided by the original manufacturers as soon as they are made available. Updates should be applied first in a controlled environment simulating the real one to minimize the effects on production~\cite{IndustrialControlSystemsCyberEmergencyResponseTeam2016RecommendedStrategies}. However, manufacturers may not offer updates for devices that have reached the end of their life-cycle, and in some cases, IIoT devices may not allow for updates or patches~\cite{EuropeanUnionAgencyforNetworkandInformationSecurityENISA2018GoodManufacturing}. In that case, compensating measures capable of reducing system vulnerabilities, such as hardening~\cite{Effendi2015ICSEnterprise}, might be studied.
   \item \textbf{Use strong passwords:} Passwords for HMIs (Human-Machine Interfaces) and workstations should be strong and unique, and they should never be the default ones~\cite{EuropeanUnionAgencyforNetworkandInformationSecurityENISA2018GoodManufacturing}. VNC (Virtual Network Computing) systems should have specific passwords for remote control. CISCO, in~\cite{Hilt2020CaughtThreats}, proves that having unprotected VNCs leaves the system vulnerable to multiple attacks. The strength of a password is related to its length, complexity and the threat model used. The human factor should also be taken into account since a lack of proper security training is a vulnerability in itself~\cite{STURM2017154}. For example, too-complex passwords may end up being written down because users keep forgetting them. Besides, under no circumstances should these passwords be related to the identity of the device they protect. Guidelines for choosing a good password are beyond the scope of this paper, but the interested reader is referred to~\cite{Fenton2017DigitalManagement}. 
   \item \textbf{Implement strict access control mechanisms:} Having some kind of access control for HMIs and workstations is strongly recommended. As was proved by~\cite{Maggi2017RogueSecurity}, doing otherwise may result in both security and safety risks. A similar approach should be considered when dealing with file servers.
   \item \textbf{Implement network segmentation:} Unrelated networks should have physical and logical separations. This is extensively explained in Section \ref{S:3.3}.
\end{itemize}

Following these recommendations enhances security by decreasing some of the most well-known vulnerabilities. However, most industrial systems require more complex security measures that fulfill the security requirements defined in the next section.\par

\subsection{Industry 4.0 Specific Security Recommendations} \label{S:2.2}
The particularities of industrial manufacturing add additional constraints in the design of efficient security approaches for OT networks. Nevertheless, the standard security requirements of IT should still be guaranteed in industrial security. They are authentication, confidentiality, access control, integrity, non-repudiation, and availability~\cite{1435744}. The following recommendations address each of them and analyze why they are important and how they can be achieved. This is done from an industrial point of view, instead of the usual IT point of view. \par
\begin{itemize}
    \item \textbf{Availability:} To guarantee this requirement, the system should be designed with fault-tolerance in mind~\cite{8463911}. Critical devices and networks should have a redundant counterpart to replace the original in the event of failure or security breach~\cite{Stouffer2015GuideSecurity}. These redundancy mechanisms help minimize the effect of DoS (Denial of Service) attacks~\cite{InternationalElectrotechnicalCommission2019IECComponents} and assure users' safety. 
    \item \textbf{Authentication and authorization:} According to the IEC 62443-4-2~\cite{InternationalElectrotechnicalCommission2019IECComponents}, every user in a system has to be authenticated, and every requester of an operation needs to be previously authorized. The advised way to achieve this~\cite{Stouffer2015GuideSecurity} is with the use of allowlists (traditionally called whitelists) and only allow communications between authenticated and authorized source-destination pairs.
    \item \textbf{Access control:} This must be considered when accessing devices' configuration and any resource in the network. Role-based access controls are strongly recommended~\cite{Stouffer2015GuideSecurity}. A strong access control system will diminish potential impersonation attacks and favor confidentiality by guaranteeing that only real users can access the system. This is of especial relevance in control systems and databases. Preventing attackers from accessing control systems prevents them from compromising industrial devices, e.g., robots as proved by CISCO in~\cite{Maggi2017RogueSecurity}. Preventing attackers from accessing databases also prevents them from getting critical information that could later be used to access critical control systems. 
    \item \textbf{Integrity and confidentiality:} Unwanted message modification can have dangerous consequences for systems and users in the IIoT. For instance, as~\cite{Tiburski2017EvaluatingE-health} presents, exposing or maliciously modifying sensitive information may put a persons' life in danger in case of a health emergency. Thus, data has to remain unchanged and confidential during capture, retrieval, update, storage, and transport. Only authorized users should be able to read or modify it. For example, as shown in Section \ref{S:6}, by using ABE only users with specific attributes or roles would be able to access the encrypted information.
    \item \textbf{Non-Repudiation:} This guarantees that messages are transmitted in a way that the authenticity of the information cannot be questioned later~\cite{El-Hajj2019ASchemes}. It is especially relevant in Human User Interfaces~\cite{InternationalElectrotechnicalCommission2019IECComponents}, so human actions are reflected in the system and can be traced back to the user who performed them.
\end{itemize}

Besides implementing the security measures mentioned above, a layered security approach is strongly encouraged. In the coming section, we introduce the concept of DiD applied to Industry 4.0 infrastructures.\par

\section{Security in Industry 4.0: A DiD approach}  \label{S:3}
One of the advanced techniques to secure industrial environments is DiD. Per the IEC 62443-4-1~\cite{InternationalElectrotechnicalCommission2018IECRequirements}, the goal of this approach is to limit the damage in case of an attack by implementing layered security controls. DiD is an effective security method that addresses many attack vectors, as each layer provides additional defense mechanisms. It can be implemented in both OT and IT networks with different security techniques but similar goals. \par

\subsection{DiD Goals} \label{S:3.1}
Most enterprises are familiar with IT security, but not so much with OT security. Until recently, it was considered that the only access points to the systems were physical and that the complexity of the industrial system itself provided protection enough~\cite{Setola2019AnSystem}. Thus, industrial security was not a concern. With the industry's evolution to Industry 4.0 and the growing connectivity of the systems, cybersecurity becomes a requirement to be implemented as part of the systems' design. Various institutions worldwide such as the NIST~\cite{Stouffer2015GuideSecurity}, the Spanish INCIBE~\cite{HerreroCollantes2015ProtocolosSCI}, and even standards as the IEC 62443-4-1~\cite{InternationalElectrotechnicalCommission2018IECRequirements} and IEC 62443-4-2~\cite{InternationalElectrotechnicalCommission2019IECComponents}  have addressed the topic of security. As~\cite{Tuptuk2018SecuritySystems} points out, this may cause a flood of information about how to integrate them in different organizations. Still, these guidelines and standards have common points that can be combined to define a DiD strategy's goals. These goals are presented below: \par

\begin{itemize}
   \item \textbf{The security requirements of Section \ref{S:2.2}}. These requirements are considered basic for any security solution with independence of the application environment. Out of all of them, availability is the main priority. Regarding data integrity, it can be compromised accidentally or as a result of an attack. The first case can be the result of interferences in industrial communications, and measures to detect unwanted modifications are already used (i.e., CRC~\cite{Koopman2015SelectionIntegrity}). However, these measures are not enough to handle active attacks. An attacker can alter the content of a control packet in a way that the CRC does not detect, and that can result in sabotage~\cite{Maxwell2003AnalysisExploits}. To protect against these attacks, a combination of role-based access control, encryption and integrity preserving algorithms (i.e., digital signatures) should be used. The access control will stop attackers from accessing the system. However, even if these are surpassed, encryption will guarantee data confidentiality (so the attacker will not be able to access the information being delivered), and the integrity preserving algorithms will guarantee that the encrypted packets cannot be altered.\par
   \item\textbf{ Restricted physical and logical access to the system}, taking into account both external and internal threats. The OT network is considered a critical network, and its connection to the IT network has to be restricted. This separation is usually achieved using a Demilitarised Zone (DMZ)~\cite{IndustrialControlSystemsCyberEmergencyResponseTeam2016RecommendedStrategies} and reducing traffic to specific and documented services and ports. The use of DMZs in combination with unidirectional gateways and firewalls restricts the logical access to the OT network and helps achieve the restricted data flow required in the IEC 62443-4-2~\cite{InternationalElectrotechnicalCommission2019IECComponents}. To restrict physical access to where critical systems are located, it is advised to use biometric systems and smart cards. The access permissions should be implemented following a least-privilege approach and issued by a trusted entity~\cite{EuropeanUnionAgencyforNetworkandInformationSecurityENISA2018GoodManufacturing}. This entity should also keep them up-to-date to reflect the current situation and prevent security breaches.
   \item \textbf{Industrial Control System (ICS) protection from known vulnerabilities.} The long lifetime of these devices makes them particularly vulnerable to attacks. When vulnerabilities are discovered, manufacturers usually offer security patches that should be installed, as explained in Section \ref{S:2.1}. In case no more security patches are available, a vulnerability assessment should be performed~\cite{DeSmit2017AnSystems}, and a rigorous hardening process should be considered~\cite{1524567}, e.g., using allowlists, reducing application services to the minimum, or restricting users' privileges and roles as much as possible.
   \item \textbf{System monitorization and security incidents detection.} Malfunctioning ICS and misconfigured services create vulnerabilities in the systems. For example, in the case of wireless devices, an incorrect configuration of security gives outsiders an access point to the industrial system~\cite{Reaves2012AnalysisSystems}. Intrusion Detection Systems (IDS) or Intrusion Protection Systems (IPS) can be implemented to detect these intrusions on time and prevent future security breaches~\cite{4660125}. IDS and IPS systems detect abnormal behaviors by comparing the current and expected status.
   \item \textbf{Periodical security evaluations.} In compliance with the guidelines of~\cite{Stouffer2015GuideSecurity}, security should be addressed during the design, use, maintenance, and removal of industrial systems. This includes hardware, software, and security policies.
   \item \textbf{Limit the impact on production.} Essential functions that guarantee health, safety, environment maintenance, and equipment availability cannot be negatively affected by security measures, or emergencies~\cite{InternationalElectrotechnicalCommission2019IECComponents}. Therefore, it is essential to find a balance that gives the system as much security as possible while still fulfilling all the production requirements. Besides, since not every attack can be prevented, industrial security frameworks should also include fast restoration plans~\cite{NIST2018FrameworkCybersecurity}.
   \item \textbf{Isolation of critical systems.} As is presented in~\cite{IndustrialControlSystemsCyberEmergencyResponseTeam2016RecommendedStrategies}
   The Internet is considered an untrusted network, so ICSs and control networks should have no connection to it. However, if this is strictly necessary, communications must use only proven secure protocols (e.g., HTTPS instead of HTTP) and go through a DMZ.
\end{itemize}

\subsection{DiD Layers} \label{S:3.2}
With the goals for a DiD strategy properly established, the next step is to choose the suitable layers. None of the standards establish which layers to use, so different approaches exist. Sometimes, instead of layers, the authors talk about elements, as in the case of~\cite{IndustrialControlSystemsCyberEmergencyResponseTeam2016RecommendedStrategies}. This approach is very helpful, as it highlights the security measures that ought to be taken. However, industrial systems would benefit from first defining the layers and then outlining the security measures within them. In order to define DiD layers appropriate for industrial systems, we compare the proposals of different authors in Table \ref{tab:4_DiDComparison}. \par

\begin{table}[!ht]
\caption{\label{tab:4_DiDComparison} Comparison between different DiD strategies.}
\setlength{\tabcolsep}{5pt}
\renewcommand{\arraystretch}{1.3}
\centering
\resizebox{110mm}{!}{%
\begin{tabular}{|c|c|l|c|}
\hline
\multicolumn{1}{|c|}{\textbf{Reference}}             & \textbf{Layers}       & \textbf{Layer Names}        
                                                           & \textbf{Designed for}     
\\ \hline
\textbf{Granzer, et al.~\cite{Granzer2011SecuritySystems}}           & 3                               & \begin{tabular}[c]{@{}l@{}}Company/Internet\\ Intranet\\ Fieldbus\end{tabular}                                                                & ICS 
\\ \hline
\textbf{Mavroeidakos, et al.~\cite{Mavroeidakos2016SecurityEnvironment}}  & 4                               & \begin{tabular}[c]{@{}l@{}}Perimeter\\ Deceptive\\ Detection\\ Cryptography\end{tabular}                                  & Cloud Computing           
\\ \hline
\textbf{Kuipers, et al.~\cite{Kuipers2006ControlStrategies}}         & 4                               & \begin{tabular}[c]{@{}l@{}}Internet and back-ups\\ Corporate\\ Control systems communications\\ Control system operations\end{tabular}    & ICS                       
\\ \hline

\textbf{Zhou, et al.~\cite{Zhou2018ConstructionSystem}}           & 5                               & \begin{tabular}[c]{@{}l@{}}Physical Protection\\ Perimeter Security\\ Intranet\\ Control System\\ Production Process\end{tabular}           & SCADA                     
\\ \hline
\textbf{Nguyen~\cite{Nguyen2017MicrosoftGuide}}             & 5                               & \begin{tabular}[c]{@{}l@{}}Network\\ Enclave boundary\\ Computing environment\\ Identity\\ Application\end{tabular}                        & Microsoft Azure           
\\ \hline
\textbf{Knapp, et al.~\cite{book:1375056}}                         & 5                               & \begin{tabular}[c]{@{}l@{}}Physical Layer\\ Network Layer\\ Application Layer\\ Data Integrity\\ Data\end{tabular}     & Generic                   
\\ \hline

\end{tabular}
}
\end{table}

As shown in Table \ref{tab:4_DiDComparison}, DiD strategies vary in their application area, the number of layers, and even the function of said layers. Most of the presented strategies work with a minimum of four layers, except for~\cite{Granzer2011SecuritySystems}, which has too few layers, so it may not be as scalable as the others. In the case of~\cite{Mavroeidakos2016SecurityEnvironment}, the presented strategy is designed for cloud computing. This proposal defines a deceptive and a detection layer. Although very interesting, matching their proposed layers with the distribution of an industrial plant may make it lose effectiveness. In contrast with this last option, the proposal of~\cite{Kuipers2006ControlStrategies} is explicitly designed for ICS. However, the proposed layers have been divided with the system architecture in mind, rather than the DiD strategy. Nevertheless, the proposed zones in the architecture are suitable, and~\cite{Zhou2018ConstructionSystem} uses them as a basis for developing their own SCADA-oriented DiD strategy. The particularity of this last approach is that it places the continued operation of the industrial system at the core of DiD strategy. Meanwhile, this paper considers that the center of the DiD strategy should be protecting the information. Therefore, this proposal solves a different situation to the one used in this article. This leads us to the last two options, out of which we discard~\cite{Nguyen2017MicrosoftGuide} for being focused solely on Microsoft Azure. The remaining proposal,~\cite{book:1375056} is based on the traditional DiD layers (Physical, Perimeter, Network, Host, Application, and Data). Since we want our DiD strategy to have information protection at its core, we will follow these traditional layers. \par

Finally, using these layers to achieve the goals presented in Section \ref{S:3.1} can be eased when applied in combination with network segmentation, first mentioned in Section \ref{S:2.1}. Segmentation is required by IEC 62443-4-2~\cite{InternationalElectrotechnicalCommission2019IECComponents} and increases security by separating the network both logically and physically.\par

\subsection{Security measures in each of the layers} \label{S:3.3}
Network segmentation enhances availability~\cite{Stouffer2015GuideSecurity} and improves the system's reliability~\cite{InternationalElectrotechnicalCommission2019IECComponents}. Segmentation can be physical or logical (e.g., gateways, firewalls, VPNs, VLANs), which might be implemented from the link layer up to the application layer. Logical segmentation is more flexible and easier to implement, but it may be bypassed and lead to single-points-of-failure, while physical segmentation is more secure, but also more complex and expensive~\cite{InternationalElectrotechnicalCommission2019IECComponents}. Thus, segmentation techniques should be analyzed case-per-case since there is no universal solution. \par
\begin{figure}[!ht]
\centering \includegraphics[width=85mm]{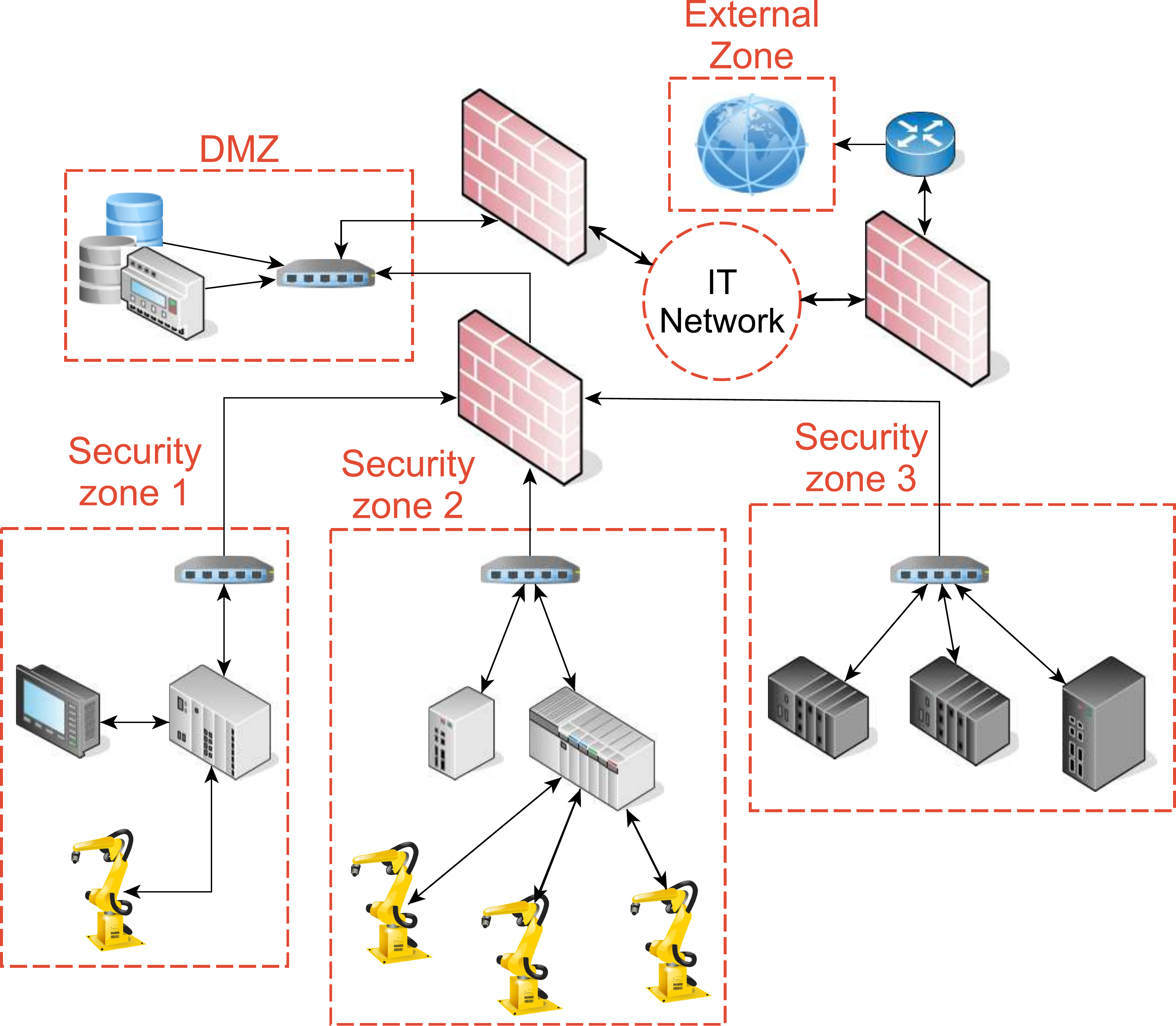}
\caption{OT network segmentation with three security zones and a DMZ separated by firewalls.}
\label{fig:FIG1}
\end{figure}
\begin{figure}[!ht]
\centering \includegraphics[width=100mm]{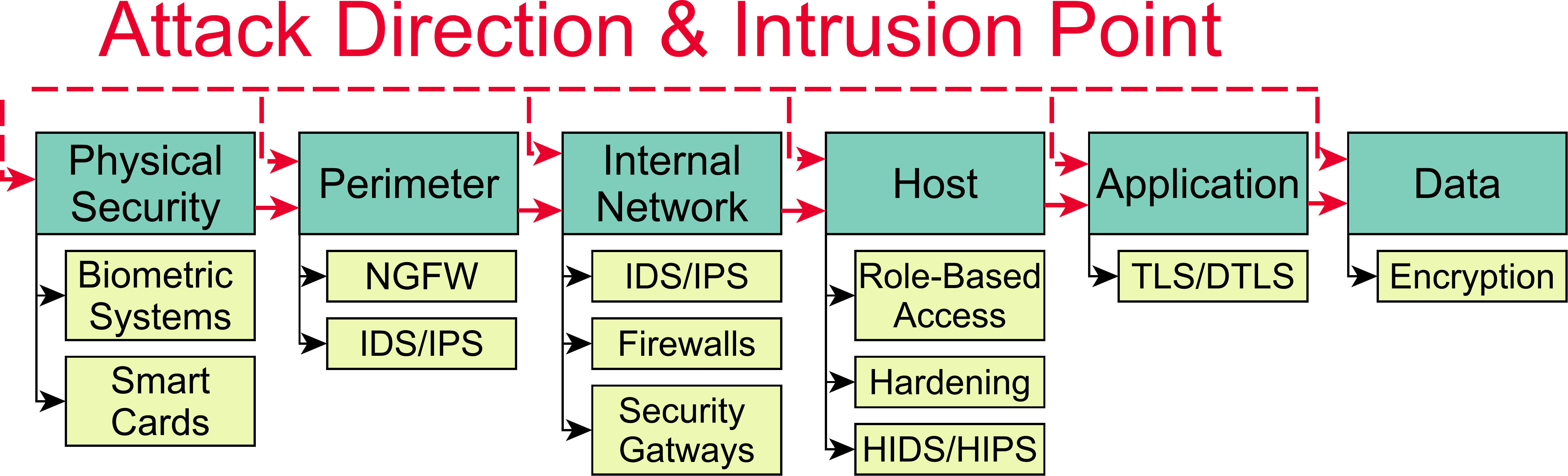}
\caption{Security layers in DiD (in blue) with their corresponding security measures (in pink).}
\label{fig:FIG2}
\end{figure}

The key to successful security frameworks lies in the combination of network segmentation (Figure \ref{fig:FIG1}) and a DiD approach. Each of the security zones should consist of assets with similar security needs~\cite{ISA2015ISA-62443-3-2:Design}, thereby facilitating monitoring and logical access control. The zones can also be subdivided into more segments as needed, improving overall security. In agreement with the IEC 62443-4-1~\cite{InternationalElectrotechnicalCommission2018IECRequirements}, the DiD layers should provide additional defense mechanisms by supporting the secure design principles specified in the same standard. The choice of which mechanisms to implement in each layer is left to the user-e.g., IDSs, IPSs, firewalls, security gateways, or encryption algorithms. Thus, following those guidelines along with the required network segmentation of the IEC 62443-4-2~\cite{InternationalElectrotechnicalCommission2019IECComponents}, a DiD layered approach is presented in Figure \ref{fig:FIG2}, where each layer has the following purposes. \par

\subsubsection{Physical Security}
The first security layer handles physical security. Measures to ensure restricted physical access must adapt to the organization's particularities. As introduced in Section \ref{S:3.1} smart cards and biometric systems are potential solutions. Smart cards contain information related to their user, which allows the subject to be identified by the card reader. More specifically, contactless smart cards are especially recommended for secure physical access in high-traffic areas since they allow for fast and accurate user identification and are almost impossible to duplicate~\cite{SmartCardAlliance2003UsingAccess}. For more critical areas, biometric systems like face or iris recognition are securer than smart cards~\cite{899930}. Unlike cards, they are also slower and more expensive, but they cannot be easily stolen or duplicated. \par
It should be taken into account that although Figure \ref{fig:FIG2} presents physical security as a single layer, this security layer is distributed through the enterprise infrastructure, and therefore it may include a wide variety of security mechanisms. For instance, access to the control room or the general assembly line may vary depending on the hour or user role. Thus, context-dependent access may be necessary. Physical security is crucial since it directly relates to the human factor. Human users can use USB sticks, cards, and other physical mediums for direct system interaction. If this access is not controlled or restricted, malicious users could use it to access critical systems and infect them~\cite{Byres2011HowSystems}. Although physical access is no longer the only access point to systems, it still has to be considered. \par

\subsubsection{Perimeter}
Perimetral security is the layer that protects the OT network from untrusted networks by restricting access and filtering unauthorized communications, including the ones coming from the IT network. Since the idea is to detect suspicious traffic, limiting it to specific ports may seem enough. However, smart manufacturing manages a high-traffic volume while the equipment may still be old. Thus, it is possible to overload the communication buffer in legacy systems and cause an accidental DoS attack~\cite{book:1375056}. Solutions based on Next-Generation Firewalls (NGFWs) should be implemented to prevent this. These firewalls can be used as shown in Figure \ref{fig:FIG1}. In it, the IT and OT networks are separated by a DMZ placed between two NGFW. As mentioned in Section \ref{S:2}, these firewalls offer deep-packet inspection and IDS/IPS functionalities, becoming very useful for network monitoring and traffic filtering tasks. Filtering is recommended to be performed following an allowlist approach~\cite{HerreroCollantes2015ProtocolosSCI}. The combination of this type of filtering with the presented paired firewall strategy can simplify firewall rules~\cite{Jiang2017ResearchSystem}. Besides, as ~\cite{Stouffer2015GuideSecurity} explains, ICS applications tend to be static, making allowlists more practical than blocklists and adding the benefit of making log analysis more manageable. \par
Regarding monitoring, it can be active or passive, depending on the particular requirements of the system. If the purpose is to analyze incidents and produce intruder alerts, IDS would be sufficient. Instead, if the aim is to stop the intrusion as soon as possible without any further analysis, IPS ought to be used~\cite{IndustrialControlSystemsCyberEmergencyResponseTeam2016RecommendedStrategies}. It is important to note that applying an IPS approach requires a deep knowledge of the network traffic since an IPS reacting to a false positive may lead to an accidental DoS~\cite{Patil2013ChapterSecurity}. Note that firewalls and IDS systems are complementary technologies, and one does not substitute the other. \par
Finally, this layer should also deal with the control of remote accesses~\cite{IndustrialControlSystemsCyberEmergencyResponseTeam2016RecommendedStrategies}. This can be done with secure VPNs, a temporal user in secured PCs, or by subjecting accessing users to vulnerability scans. \par

\subsubsection{Internal Network}
So far, the proposed security layers protect the system as a whole and are designed to avoid unauthorized access from outside the trusted network. In contrast, the following security layers are devised to protect network resources when attackers are already within the network, which, according to the summary provided in~\cite{Wu2018CybersecurityManufacturing}, is one of the vulnerabilities of digital manufacturing systems.\par 
Therefore, the security measures in this layer will be applied independently to each of the security zones or sub-networks. The security measures of this layer are mainly composed of devices that control the sub-network inbound, and outbound traffic~\cite{book:1375056}, such as IDSs/IPSs, firewalls, and security gateways. \par
The sophistication of cyberattacks is continually growing, so stopping them keeps requiring more complex security measures~\cite{Jang-Jaccard2014ACybersecurity}. Applying the sophisticated security measures to smaller networks improves their efficiency and allows them to be specifically designed with the sub-network requirements in mind.\par

\subsubsection{Host}
The goal of the next layer is to protect each of the devices inside a security zone. This is of particular relevance in OT security, where targeted attacks on critical systems may cause significant damage to the whole system~\cite{10.1145/2667190.2667192}. For example, an attacker can modify a device's firmware to gain control of it~\cite{7749239}. Thus, it is crucial to detect anomalies by actively scanning for vulnerabilities and modifications in the firmware or device configuration. The security measures applied in this layer vary depending on the system's capabilities and limitations. If newer devices support role-based access control, it is advisable to apply it~\cite{Stouffer2015GuideSecurity}. This measure can be reinforced by following the recommended practices in Section \ref{S:2.1} and the hardening practices introduced in Section \ref{S:3.1}. In the case of legacy devices that cannot implement advanced authentication mechanisms~\cite{EuropeanUnionAgencyforNetworkandInformationSecurityENISA2019INDUSTRYRECOMMENDATIONS}, reinforced access control to the security network they are located in should be considered. Finally, additional security measures at the host level can also be considered if the asset supports them, such as Host-Based IDS (HIDS) or Host-Based IPS (HIPS). These would provide another layer for monitoring and detecting abnormal situations in the host. \par

\subsubsection{Application and Data}
These layers are the last safeguards against attacks and are directly related to IT security. They aim to protect data and services from attacks that the previous layers have not detected. It is strongly recommended to secure the communications between applications with protocols like TLS or DTLS, combined with data encryption~\cite{Heer2016Secure4.0}. Even if they remain independent, these layers are closely related since one may directly affect the other. This will be further explained in Section \ref{S:4}. These layers deal with the worst-case scenario: an attacker that has infiltrated the system and can directly interact with the information generated in it. Thus, the main goals of these layers will be protecting data confidentiality and integrity.\par
The proposed DiD layers fulfill the requirements of Section \ref{S:3.1}, as shown in Table \ref{tab:5_CruceReq}, and accomplish all the goals of a DiD strategy, some even in more than one layer. Despite this redundancy, the IEC 62443-4-1 DiD recommendations are fulfilled since the layers remain autonomous and similar functionalities are achieved by different means. Thus, if an attacker breaks into the system, they still have to surpass many security barriers with different weaknesses before achieving their goal. \par

\begin{table}[!ht]
\caption{\label{tab:5_CruceReq} Goals covered by the proposed security layers. \Circle No ;  \CIRCLE Yes; \LEFTcircle Some cases}
\renewcommand{\arraystretch}{1.2}
\centering
\resizebox{100mm}{!}{%
    \begin{tabular}{cc|c|c|c|c|c|c|}
    \cline{3-8}
    \textbf{}                                                                            &                                 & \rotatebox{90}{\textbf{\begin{tabular}[c]{@{}c@{}}Physical \\ Layer\end{tabular}}}                           & \rotatebox{90}{\textbf{Perimeter}}                               & \rotatebox{90}{\textbf{\begin{tabular}[c]{@{}c@{}}Internal \\ Network\end{tabular}}}                         & \rotatebox{90}{\textbf{Host}}                                 & \rotatebox{90}{\textbf{Application }}                              & \rotatebox{90}{\textbf{Data}}                                     \\ \hline
    \multicolumn{2}{|c|}{\textbf{Restricting Physical Access}}                                                                 & \CIRCLE                      & \Circle                     & \Circle                     & \Circle                      & \Circle                     & \Circle                     \\ \hline
    \multicolumn{1}{|c|}{\multirow{2}{*}{\textbf{\begin{tabular}[c]{@{}c@{}}Restricting\\ logical access\end{tabular}}}}               & \textit{To Network}             & \Circle                     & \CIRCLE                      & \CIRCLE                      & \Circle                      & \Circle                     & \Circle                     \\ \cline{2-8} 
    \multicolumn{1}{|c|}{}                                                                   & \textit{To Devices}             & \Circle                     & \Circle                     & \Circle                     & \CIRCLE                       & \Circle                     & \Circle                     \\ \hline
    \multicolumn{2}{|c|}{\textbf{Hardening}}                                                                                   & \Circle                     & \Circle                     & \Circle                     & \CIRCLE                       & \Circle                     & \Circle                     \\ \hline
    \multicolumn{1}{|c|}{\multirow{2}{*}{\textbf{\begin{tabular}[c]{@{}c@{}}Protecting unwanted \\ modification of data\end{tabular}}}} & \textit{Role-Based Access}               & \CIRCLE                     & \Circle                     & \Circle                     & \CIRCLE                       & \Circle                     & \LEFTcircle                      \\ \cline{2-8} 
    \multicolumn{1}{|c|}{}                                                                   & \textit{Encryption} & \Circle & \Circle & \Circle & \Circle & \CIRCLE & \CIRCLE \\ \hline
    \multicolumn{2}{|c|}{\textbf{Monitoring}}                                                                                  & \Circle                     & \CIRCLE                      & \CIRCLE                      & \CIRCLE                       & \Circle                     & \Circle                     \\ \hline
    \end{tabular}
}
\end{table}

In summary, Industry 4.0 requires that IT and OT work together from the design stage on behalf of network security. For this purpose, passive mechanisms such as access control, traffic analysis, and intrusion detection should be combined with active mechanisms like traffic filtering, vulnerability scanning, and hardening. It is also of the utmost importance to provide the information collected throughout all these layers, clearly and comprehensively, to deal with potential problems as soon as possible. Finally, all of these mechanisms must be applied considering network segmentation. Every middlebox or node used to connect assets is likely to have full access to data, so E2E security measures ought to be studied and implemented. \par

\section{Encryption for Industry 4.0} \label{S:4}
Industry 4.0 handles sensitive information related to the manufacturing process. Therefore, maintaining data confidentiality is vital to any Industry 4.0 security architecture, which is achieved through cryptography. However, IIoT devices (e.g., smart robots, gateways, sensors, or actuators) are heterogeneous in terms of memory, communication, and processing capabilities. These constraints must be considered since encryption and decryption are computationally expensive operations and may introduce latencies. Initially devised for the IoT, lightweight encryption ciphers may be suitable for the IIoT. As was introduced in~\cite{Tuptuk2018SecuritySystems} IoT security techniques may be applied to smart manufacturing, as long as the particularities of the new domain are addressed. Thus, although there are challenges to applying encryption in industry, there are also mechanisms to reduce its impact as long as network security requirements and computing limitations are assessed. For instance, asymmetric cryptography requires a high amount of computing and memory resources compared to symmetric cryptography, and it is best suited for administrative purposes~\cite{Stouffer2015GuideSecurity}. Meanwhile, symmetric cryptography can be applied to the data stream, and network traffic~\cite{Stouffer2015GuideSecurity}, but it involves sharing a key beforehand, and this is not always possible~\cite{Sisinni2018IndustrialDirections}. Finally, it is essential to note that while some IIoT nodes will perform state-of-the-art encryption, others may not have the processing power for it. In this case, relegating cryptography to hardware accelerators~\cite{Stouffer2015GuideSecurity} may be the only available solution. In any case, encryption is encouraged to be included in the design of E2E security architectures whenever possible, especially in wireless networks. \par
Intending to develop encryption schemes suitable for industrial environments, researchers have studied different options to achieve this. Authors in~\cite{Fauri2017EncryptionCurse} suggest using TLS to protect the communications between systems, which, as was introduced in Section \ref{S:1}, does not provide E2E security in the presence of compromised middleboxes. The other possibility to protect the exchanged messages is the use of encryption algorithms, like the one in~\cite{10.1145/3174776.3174777}. The authors use an Open Source PLC to embed the encryption in the controller itself. However, most PLCs in industrial environments are not open source, so this solution's applicability is greatly reduced in real environments. Researchers in~\cite{6172112},~\cite{10.1007/978-3-642-10847-1_36} and~\cite{10.1007/978-3-642-23948-9_4} analyse the possibility of using symmetric and asymmetric encryption in SCADA. However, as has been mentioned, asymmetric cryptography is too heavy for industrial systems, and symmetric encryption requires exchanging keys beforehand. The key exchange and management systems, as~\cite{Tuptuk2018SecuritySystems} explains, tend to require too much computational power and may interfere with communication times. A similar conclusion about not having an industrial-appropriate key management system is expressed by~\cite{4511554}. Authors in~\cite{Kang2011ProposalSystems} also study symmetric encryption and key management but do not consider the existence of middleboxes between the communicating devices. Therefore, there is still the need for an encryption scheme compatible with the constrained nature of industrial devices, which includes a key management solution and provides the system with E2E security. \par

\subsection{Towards E2E Security} \label{S:4.1}
\begin{figure}[!ht]
\centering
\includegraphics[width=85mm]{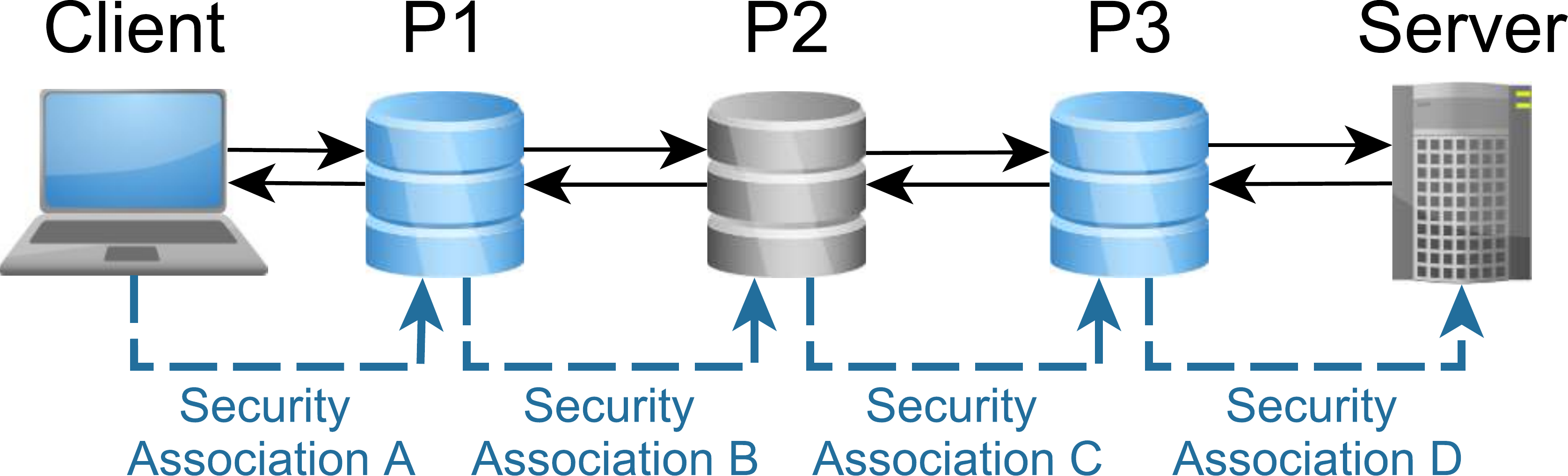}
\caption{Hop-By-Hop Security. Security is guaranteed for every security association, but not from Client to Server.}
\label{fig:FIG3}
\end{figure}

\begin{figure}[!ht]
\centering
\includegraphics[width=85mm]{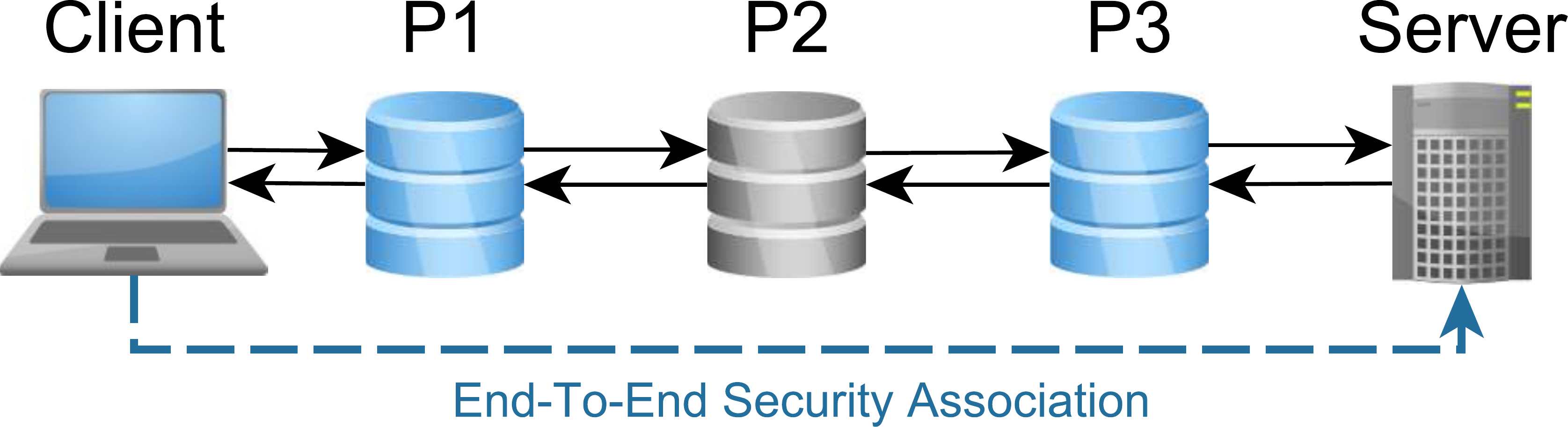}
\caption{E2E Security. Middleboxes only have access to the information they need to forward the message to the next endpoint.}
\label{fig:FIG4}
\end{figure}

Section \ref{S:3.3} shows the need to introduce intermediate entities (like gateways and proxies) to achieve security in network segmentation. IIoT devices may use lightweight communication protocols, such as MQTT~\cite{Banks2019MQTT5.0} or AMQP~\cite{OASIS2012OASIS1.0}, and these need to be translated to industrial communication protocols (e.g., Profibus, Profinet, Ethernet/IP, or EtherCAT). Protocol translation happens in gateways that need access to the data, so messages must be constantly decrypted and encrypted again, breaking security at every middlebox (Figure \ref{fig:FIG3}). Thus, instead of E2E security (i.e., secure communication is guaranteed from the sender to the final destination, Figure \ref{fig:FIG4}), there is hop-by-hop security, which does not maintain the required confidentiality if the intermediate entities are compromised.\par

E2E security requires maintaining confidentiality and integrity up to the destination while allowing proxies and gateways to do their jobs. For this to happen, these devices should only have access to the indispensable parts of the message, while the rest is hidden from them. Typically, asymmetric and symmetric encryption schemes view encryption as an all-or-nothing operation (i.e., the user either decrypts the entire message or learns nothing about it~\cite{boneh2011functional}). Thus, middleboxes would get too much information, making these ciphers not the best suited for decentralized architectures. As such, it might be necessary to encrypt data to be shared at a fine-grained level. This can be achieved with object security~\cite{mattsson2014object}, which would encrypt the payload while leaving the header unencrypted. \par
An application layer protocol that uses object security is
OSCORE (Object Security for Constrained RESTful Environments)~\cite{rfc8613}. It uses EDHOC (Ephemeral Diffie-Hellman Over COSE) to exchange keys and  COSE (CBOR Object Signing and Encryption)~\cite{rfc8152} for security. Because of the optimization of these protocols for constrained environments, this paper focuses on their combined use as the potential object security solutions for Industry 4.0. \par
Another aspect to be addressed in E2E security is the possibility of parties outside the OT network having to access the data generated in it. This data retrieval will occur in the DMZ, as explained in Section \ref{S:3}, while confidentiality still must be preserved. To this end, it would prove useful to have an encryption mechanism that enables multiple users to access the information without re-encrypting it repeatedly or distributing new keys. This can be accomplished with Functional Encryption~\cite{boneh2011functional}—i.e., IBE (Identity-Based Encryption)~\cite{Boneh2003Identity-BasedPairing} and ABE~\cite{sahai2005fuzzy}. These ciphers encrypt information according to a set of identities (IBE) or attributes (ABE) that users must possess if they want to decrypt it. ABE can therefore be considered an evolution of IBE since it provides more flexibility by encrypting data in a more detailed manner. This article will cover ABE since attributes provide a more flexible way of defining who can read encrypted data. \par
Summarising, OT networks require efficient lightweight communication and encryption protocols. In this context, object encryption combined with lightweight data formats balance security and computational cost and can be integrated into the Application and Data layers of the proposed DiD strategy. Section \ref{S:5} focuses on this possibility. Meanwhile, Section \ref{S:6} presents a detailed description of attribute-based encryption, which provides role-based access to ciphertexts. This allows them to be shared with different endpoints without the user who encrypts data identifying them one by one while guaranteeing data confidentiality. \par

\section{Object Security} \label{S:5}
Object security aims to protect the message itself, providing fine-grain access control of its content. This is achieved using ``Secure Objects", which are information containers consisting of a header, an encrypted payload, and an integrity verification tag~\cite{mattsson2014object}. The same message may carry several objects, or different parts of the message can be individually protected. Thanks to this property, object security is an effective way to obtain E2E security through middleboxes since messages can be encrypted so that middleboxes can only read the required information. Therefore, even if intermediate nodes are compromised, payload confidentiality is not jeopardized. The object security method for constrained environments proposed by the IETF Working Group, CoRE, is OSCORE. It uses the CBOR data format, COSE for encryption, and EDHOC as the key management protocol. They are explained in the following sections. \par

\subsection{CBOR} \label{S:5.1}
The need for an object data format for constrained devices arose with the presentation of the Object Security Architecture for the IoT (OSCAR)~\cite{Vucinic2015OSCAR:Things}. This architecture had low energy consumption, low latency and ensured security through middleboxes. However, it did not include an object security format suitable for constrained devices, so the architecture's efficiency was reduced in such scenarios~\cite{mattsson2014object}. To solve this, the IETF proposed CBOR~\cite{rfc7049}, a data format optimized for highly constrained environments. It uses a binary type data format, which reduces human-readability, but increases the message transmission and coding/decoding speeds.\par

\subsection{COSE} \label{S:5.2}
COSE~\cite{rfc8152} was proposed to provide CBOR with security mechanisms, such as the creation and processing of signatures, message authentication codes, and encryption. It specifies which signature algorithms shall be applied and how to build, encrypt and decrypt messages. COSE messages are constructed in ``layers", allowing for the sought fine-grain-level approach. The standard offers different encryption and signing possibilities, but when working with OSCORE, it only uses the untagged COSE\textunderscore Encrypt0 structure. \par 
This protocol does not specify the message's recipients and assumes that they know the key to be used for decryption. Therefore, it should be combined with key management protocols like EDHOC. \par

\subsection{EDHOC} \label{S:5.3}
EDHOC is a lightweight key exchange protocol with a small message overhead~\cite{ietf-lake-edhoc-01}, making it efficient for technologies with duty-cycle or battery limitation. According to the standard, EDHOC also provides the following security features: \par
\begin{itemize}
  \item \textbf{Mutual authentication with aliveness.} This means that the communicating parts authenticate each other. This way, both endpoints know they are communicating with whom they intended. It helps reduce impersonation attacks.
  \item \textbf{Perfect Forward Secrecy (PFS).} EDHOC achieves this by running an Elliptic Curve Diffie-Hellman (ECDH) key exchange with ephemeral keys. It guarantees that if an attacker gets the keys, it only gets the ones being used at the moment of an attack, and every message exchanged with previous keys continues to be confidential. 
  \item \textbf{Identity protection.} Passive attackers cannot learn the identity of either communicating party. Active attackers can only learn about the receiver~\cite{Bruni2018FormalEDHOC}.  
  \item \textbf{Crypto Agility, given by COSE.} This facilitates changing the cryptography algorithms, making potential system upgrades faster and easier.
  \item \textbf{Protection against replay attacks.} This prevents attackers from re-sending messages that have already been received. 
  \item \textbf{Protection against message injection.} This prevents an attacker from injecting fake messages into the stream.
\end{itemize}

Although EDHOC does not add requirements to the transport layer, it is recommended to implement it in combination with CoAP (Constrained Application Protocol)~\cite{rfc7252}, CoRE's communication protocol for constrained devices. They have also developed a draft with new configuration options to improve CoAP default security, including the prevention of amplification attacks. Its implementation is encouraged to prevent IIoT devices from being manipulated to launch DDoS attacks. The interested reader is referred to~\cite{ietf-core-echo-request-tag-10} for more details about these enhancements.\par
\begin{figure}[!ht]
\centering
\includegraphics[width=73mm]{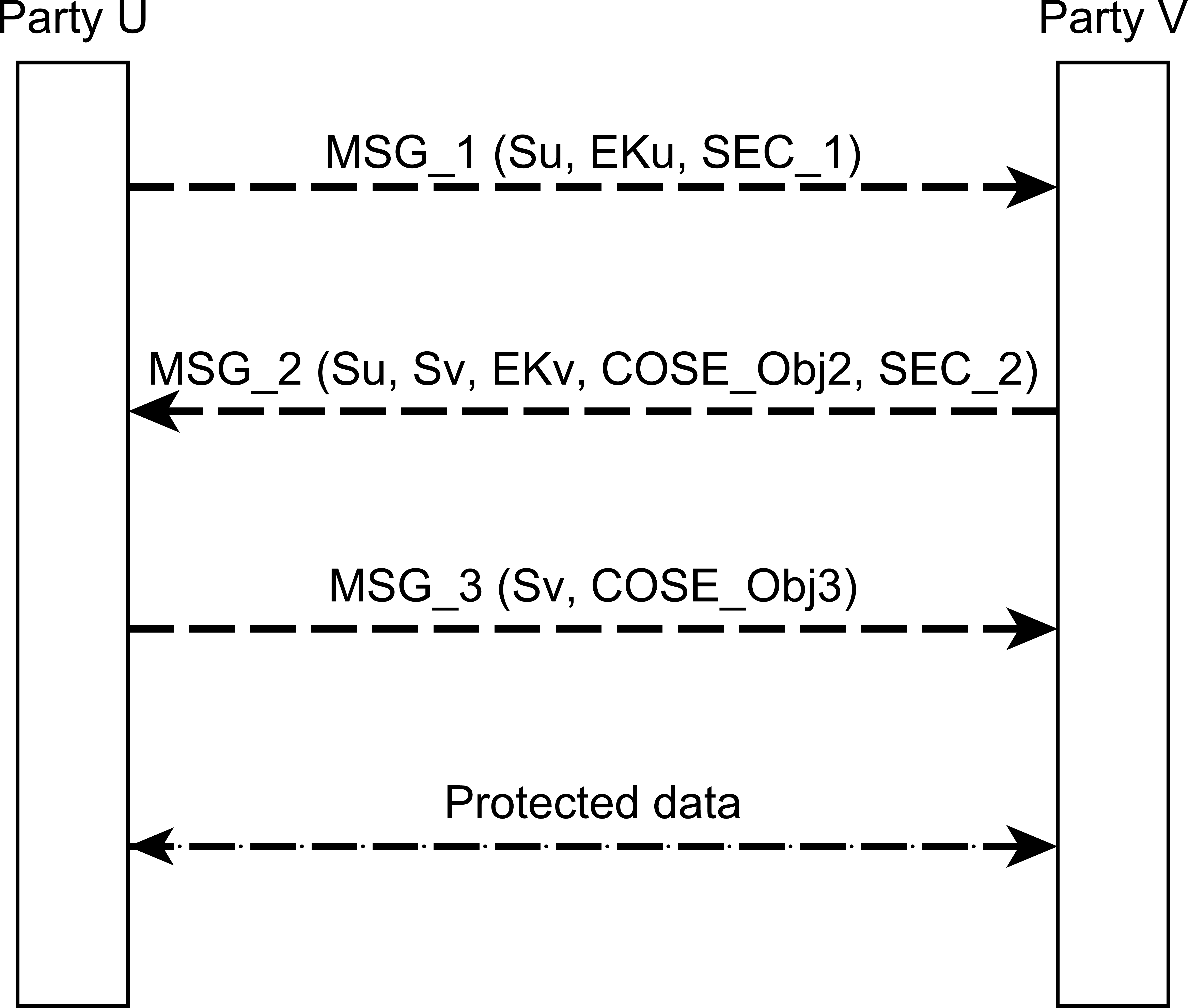}
\caption{EDHOC negotiation messages.}
\label{fig:FIG5}
\end{figure}

EDHOC key exchange takes three messages between a Party U (initiator) and a Party V (responder), after which message exchange between both parties is protected. Each of these three messages is a CBOR sequence protected by COSE. EDHOC supports various authentication methods—i.e., certificates, PSK (pre-shared key), and RPK (raw public key). The parameters exchanged between parties will vary between methods, but a simplification is included in Figure \ref{fig:FIG5}. \par

In Figure \ref{fig:FIG5}, MSG\textunderscore 1 includes party U's session key (Su) and ephemeral key (EKu), and SEC\textunderscore 1. SEC\textunderscore 1 specifies the supported elliptic curves for the ECDH and the supported cipher suites. MSG\textunderscore 2 answers with both party's session keys (Su and Sv), V's ephemeral key (EKv), COSE\textunderscore Obj2 and SEC\textunderscore 2. SEC\textunderscore 2 now contains the selected elliptic curves and cipher suites. Finally, MSG3 contains Party V's session key and COSE\textunderscore Obj3. As it is summarised in~\cite{Perez2019ApplicationIoT}, COSE\textunderscore Obj2 is used to protect MSG\textunderscore 1 and MSG\textunderscore 2 integrity and to authenticate the server. Meanwhile, COSE\textunderscore Obj3 authenticates the client and ensures the integrity of the exchanged messages. \par

The security features of EDHOC are in line with the security requirements for Industry 4.0 detailed in Section \ref{S:2}. For instance, the protection against replay and message injection attacks may prevent an attacker from sabotaging the control messages. Moreover, since it provides PFS, EDHOC helps mitigate pervasive monitoring, preventing an attacker from learning more about the system to prepare for a more harmful attack. Finally, the first message exchanged in EDHOC allows verifying that the chosen cipher suite is supported by both communicating parties, which is necessary for the commonly heterogeneous manufacturing environments. \par

\subsection{OSCORE} \label{Sec_ObjectSecurity_OSCORE}
OSCORE~\cite{rfc8613} is CoRE's application layer security framework for constrained environments. It uses EDHOC as a key exchange protocol and protects messages using COSE. Integrity and confidentiality are provided by the Authenticated Encryption with Associated Data algorithm (AEAD)~\cite{rfc5116}, while authentication and authorization come from using the Authentication and Authorisation for Constrained Environments (ACE) standard~\cite{ietf-ace-oauth-authz-35}. \par
\begin{figure}[!ht]
\centering
\includegraphics[width=100mm]{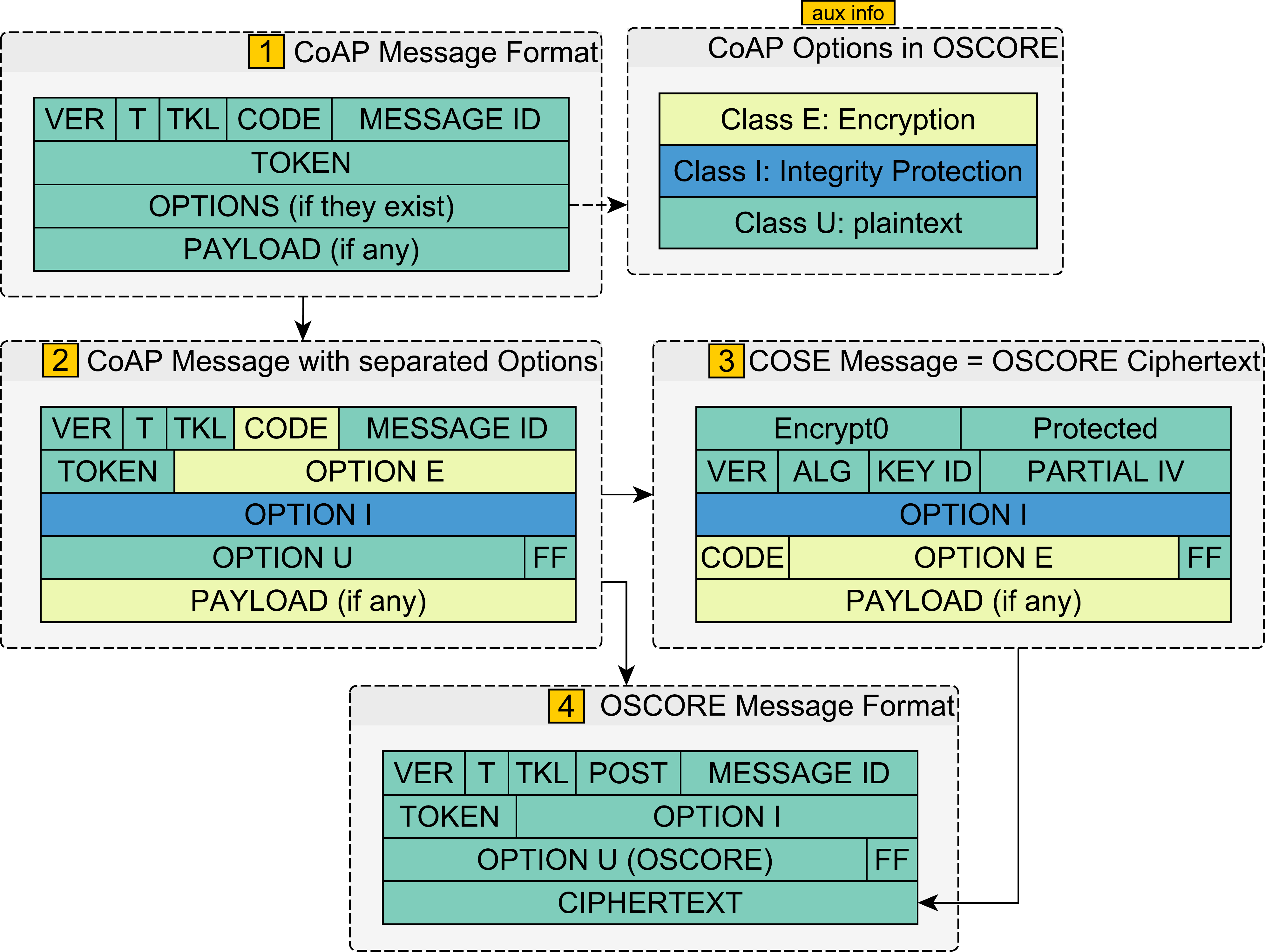}
\caption{Composition of OSCORE messages.}
\label{fig:FIG6}
\end{figure}

OSCORE also improves COSE's security by encrypting the method in the original header and placing it in the encrypted payload. A dummy code is then placed in the new header: POST for requests and CHANGED for responses. This prevents attackers from changing a PUT to a DELETE and deleting a resource. Figure \ref{fig:FIG6} shows how OSCORE messages are built upon CoAP messages. Some fields are encrypted, others only integrity protected, and others are left in plaintext (box 2). This information is encapsulated in a COSE message (box 3), which is the content of the ciphertext field of the OSCORE message (box 4). Therefore, the payload is now encrypted, while the header fields remain in plain text and can be processed by middleboxes, if necessary. \par

Apart from providing E2E security even in the presence of middleboxes, OSCORE guarantees most of the industrial security requirements specified in Section \ref{S:2.2}. These include integrity, authentication, and authorization. Moreover, OSCORE is specially designed for constrained networks, making it highly optimised for IIoT nodes. As shown in~\cite{gundougan2020iot}, it has less overhead than CoAP+DTLS, it is faster both in single-hop and multiple-hop scenarios, and it also deals better with retransmissions. Finally, the combined use of OSCORE and EDHOC has a small footprint~\cite{ietf-lake-edhoc-01}, thanks to the fact that both use CBOR and COSE. \par 
The use of these protocols, specially designed for constrained devices, make OSCORE very useful for securing messages between the IIoT nodes constituting an OT network. Furthermore, EDHOC provides the PFS OSCORE cannot provide by itself. In case the keys are compromised, this property ensures that every encrypted message exchanged in previous sessions remains protected. 
Industry 4.0 will also benefit from OSCORE's header compression and it being mappable to HTTP. The compression reduces the per-packet overhead, making the transmission of industrial small data packets faster. The compatibility with HTTP facilitates the connectivity IIoT nodes need. \par

\section{Attribute-Based Encryption} \label{S:6}
As shown in Section \ref{S:5}, OSCORE protects requests and responses using partially encrypted messages. It also uses CoAP as the communication protocol, which supports requests to an IP multicast group~\cite{rfc7390}. However, protecting group messages with OSCORE~\cite{ietf-core-oscore-groupcomm-09} entails challenges such as handling, distributing, and updating keys. As a result, the efficiency of OSCORE is reduced in situations where data needs to be encrypted and distributed to a group whose members change frequently. ABE can solve this issue by relating ciphertexts to attributes. In Industry 4.0, it may be applied to confidential or sensitive information that has to be accessed by parties from outside the OT network. This can be the case of audit logs~\cite{goyal2006attribute}: each entry could be encrypted according to an access policy, giving different endpoints particular access rights to the same bulk of data without worrying about key distribution. \par

Because ABE creates ciphertexts according to a set of attributes or roles, senders do not need to know the identity of every recipient. This allows data to be encrypted once and shared with multiple users, simplifying key management compared to OSCORE. For instance, in a publisher-subscriber communication model (e.g., MQTT, AMQP or CoAP Pub/Sub~\cite{ietf-core-coap-pubsub-09}), the use of ABE means that the group key does not have to be updated or the information re-encrypted whenever a new node joins the network, improving scalability~\cite{Wang2014PerformanceIoT}. This makes ABE a very interesting encryption mechanism for Industry 4.0. \par 

In ABE a user with a private key \textomega\ may decrypt data encrypted with the public key \textomega ', if and only if the difference between \textomega\ and \textomega '\ is minimal~\cite{sahai2005fuzzy}. What constitutes these keys depends on whether the chosen approach is Key-Policy ABE (KP-ABE)~\cite{goyal2006attribute} or Ciphertext-Policy ABE (CP-ABE)~\cite{Bethencourt2007Ciphertext-policyEncryption}. In KP-ABE the plaintext is encrypted according to a subset of attributes. Meanwhile, in CP-ABE the plaintext is encrypted according to a policy that dictates which attributes must be fulfilled to decrypt the message. CP-ABE is more interesting for Industry 4.0 applications because it gives the sender of the message complete control over who will be capable of decrypting it. This is called implicit authorization, and it works as follows:
\begin{enumerate}
    \item Private keys are associated with an arbitrary number of attributes expressed as strings. For example: \par
    - A database in Secure Zone A has the attributes: $\{ \textrm{``Zone A"} \wedge \textrm{``Database"} \}$.\par
    - A robotic cell in Secure Zone A has the attributes: $\{ \textrm{``Zone A"} \wedge \textrm{``RobCell"}\}$.\par
    - A database in Secure Zone B has the attributes: $\{\textrm{``Zone B"} \wedge \textrm{``Database"} \}$.\par
    \item The ciphertext specifies an access policy/structure over a defined universe of attributes within the system. The sender establishes the policy. For example, a temperature sensor sends readings with the following access structures: \par
    - Temp. 01: $\{\textrm{``Zone A"} \wedge \textrm{ (``Database"} \vee \textrm{``RobCell"})\}$ \par
    - Temp. 02: $\{\textrm{``Zone A"} \wedge \textrm{``Database"}\}$ \par
    \item The recipient may decrypt the ciphertext if and only if its attributes fulfil the ciphertext's access structure.\par
    In this case, the database in Secure Zone A can decrypt both temperatures, the robotic cell is only able to decrypt the first one, and the database in Secure Zone B can decrypt neither. \par
\end{enumerate}

In an Industrial environment, ABE achieves E2E security and provides role-based access control to data. In Industry 4.0, it is becoming more usual for entities outside the OT network to need access to the data generated in it. The privileges of these entities have to be controlled and limited according to their needs. Using ABE over CoAP to encrypt the information provided to these entities ensures that only legitimate endpoints can decrypt it. \par

Finally, integrating ABE in a DiD framework should be straightforward. DiD calls for role-based access whenever possible, and thus the structures to define the access policies should already be in place. Therefore, these trusted entities can also be used to distribute the original attributes of ABE. \par

\section{CONCLUSIONS} \label{S:7}
This paper presents an overview of security measures and recommendations for a secure Industry 4.0. Along with this, it also presents an analysis on how to achieve E2E security in industry, which is usually not guaranteed in the presence of some intermediate elements, such as proxies or gateways.\par

First, best practices to secure Industry 4.0 are identified. These aim to enhance OT network security by adapting and implementing IT security recommendations to Industry 4.0. They cope with authentication, confidentiality, integrity, availability, and non-repudiation. However, most Industry 4.0 environments will require more sophisticated implementations to meet those requirements. For this reason, a DiD approach is suggested. \par

In a DiD strategy, security is divided into layers to address as many attack vectors as possible. These layers can be adapted to company criteria, but they should guarantee the following: restricted access to the network and IIoT devices, the separation of OT and IT networks as well as critical systems and to protect the ICS from vulnerabilities by installing security patches and using security measures that protect data confidentiality, such as encryption. Compliance with these requirements should be reviewed in the periodical evaluations of security and be accompanied by corporate policies that ensure rapid system restoration. After studying the literature, we have chosen to use the traditional DiD layers: Physical, Perimeter, Network, Host, Application, and Data. These layers are presented along with the technologies considered for them. Once the techniques to be applied in each layer have been defined, compliance with the proposed security goals for DiD is verified. Among the technologies and procedures studied, role-based access control combined with the principle of least privilege is strongly recommended. Additionally, the use of NGFW and DMZ has been proposed to segregate IT and OT. It is also suggested to combine these firewalls with IDS and IPS to monitor inbound and outbound traffic while highlighting the importance of avoiding false positives from IPS. Finally, keeping sensitive information confidential is vital in Industry 4.0, so the integration of E2E encryption in the DiD architecture is analyzed. \par

The proposed solutions for E2E security are OSCORE and ABE. OSCORE provides E2E security by encrypting the message payload and leaving the header fields in plaintext. Thus, gateways can process messages without breaking their confidentiality. OSCORE is concluded to be an appropriate security framework for Industry 4.0 thanks to its header compression, data format, and optimized key exchange protocol. Another feature that reinforces this conclusion is its capability of working with HTTP, which enhances the connectivity of IIoT devices. ABE is the encryption proposed to manage third-party access to the information contained in the OT network. Since IIoT nodes are highly structured and changes are rare and predictable, any outsider temporarily accessing the system is considered a vulnerability in the Industry 4.0 security framework. To counter this, we propose to encrypt the data required by these parties with ABE. This allows fine-grained access control to sensitive data and simplifies key management, avoiding having to issue new keys and to re-encrypt messages whenever a new entity accesses the system. Besides, ABE is determined to have easy integration into the DiD environment. The trusted third-party used to define the roles for role-based access can be employed to determine and distribute the attributes and the access policies for the information to be shared. \par

\section*{Acknowledgment}
This work was financially supported by European commission through ECSEL-JU 2018 program under the COMP4DRONES project (grant agreement N$^{\circ}$ 826610), with national financing from France, Spain, Italy, Netherlands, Austria, Czech, Belgium and Latvia. It was also partially supported by the Ayudas Cervera para Centros Tecnológicos grant of the Spanish Centre for the Development of Industrial Technology (CDTI) under the project EGIDA (CER-20191012), and by the Basque Country Government under the ELKARTEK program, project TRUSTIND - Creating Trust in the Industrial Digital Transformation (KK-2020/00054).

\bibliographystyle{splncs04}
\bibliography{NEWreferences}

\end{document}